\documentclass[twocolumn,showpacs,preprintnumbers,amsmath,amssymb,floatfix]{revtex4}

\usepackage{graphicx}
\usepackage{dcolumn}
\usepackage{bm}
\usepackage{float}

\input{psfig.sty}

\begin{document}

\title{Dynamics of the excitations of a quantum dot in a microcavity}

\author{J.I. Perea}

\author{D. Porras}
\altaffiliation[Present address: ]
{Max-Planck Institute for
Quantum Optics, D-85748, Garching, Germany.}

\author{C. Tejedor}
\affiliation{Departamento de F\'{\i}sica Te\'orica de la Materia Condensada,
Universidad Aut\'onoma de Madrid, 28049 Madrid, Spain.}

\begin{abstract}
We study the dynamics of a quantum dot embedded in a three-dimensional
microcavity in the strong coupling regime in which the quantum dot exciton has an
energy close to the frequency of a confined cavity mode. 
Under the continuous pumping of the system,
confined electron and hole can recombine either by
spontaneous emission through a leaky mode or by stimulated emission of a cavity
mode that can escape from the cavity. The numerical integration of a master
equation including all these effects gives the dynamics
of the density matrix. By using the quantum regression theorem, we compute
the first and second order coherence functions required to
calculate the photon statistics and the spectrum of the emitted light. 
Our main result is the determination
of a range of parameters in which a state of cavity modes with poissonian
or sub-poissonian (non-classical) statistics can be built up within the microcavity. 
Depending on the relative values of pumping and rate of stimulated emission,
either one or two peaks close to the excitation energy of the dot and/or
to the natural frequency of the cavity are observed in the emission spectrum. 
The physics behind these results is discussed.

\end{abstract}

\maketitle

\section{Introduction}
Quantum electrodynamics of atoms within optical cavities is a well understood
problem which has produced many results of both fundamental and practical
interest\cite{cohen,walls,scully,yamamoto}. The current capability of 
using semiconductors technology to grow quantum dots (QD)
embedded in microcavities seems very promising to reproduce and control that
properties in a solid-state system that could be integrated in
electronic or optical devices. The essential physics of such system is a
strong coupling between the QD excitations
and the photonic cavity modes as well as the possibility of interaction with
the external world through the pumping of the system and the subsequent
emission of light. 
The main goal is the control of the emitted light
and its quantum properties being possible to built up high efficiency light
emitting diodes, low threshold lasers and single photon sources.

We concentrate here in a problem, no far from the actual situation in several
experiments \cite{gerard,benson,michler,solomon,moreau,santori,pelton},
in which a single QD is embedded in a
pillar or disk microcavity confining photons in the three spatial directions.
The system is pumped either by electrically injecting electrons and holes
which tunnel into the QD\cite{benson},
or by pumping excitons in the QD \cite{gerard,michler,moreau,santori,pelton}.
Light emission form the system takes place both by cavity mode decay or by
spontaneous emission of leaky modes. All these pumping and emission mechanisms
introduce decoherence affecting the quantum properties of the system.

The lowest energy excitation of a neutral QD is an electron-hole pair, usually
labelled as exciton. Due to the fermionic character of its components,
this exciton state can only be singly occupied having a degeneracy related to
the possible values of the third component of the total angular momentum. This
quantum number is usually referred to as the spin of the exciton. Among all
these states, only two cases, those corresponding to $\pm 1$, are of interest
in our problem due to their possible coupling with photons. The next
excitation corresponds to the case in which both $+1$ and $-1$ excitons are
occupied. This biexciton state has an energy different to twice that of a single
exciton due to the Coulomb interaction of their components. This spectrum presents
interesting alternatives\cite{stace} that we intend to study in future work
while here we restrict to simplest case, in which we consider only excitons with 
a given spin. This assumption is well justified in the case of experiments in which
the system is pumped with polarized light, so that excitons with a given 
angular momentum are created in the system. We also assume that spin
flip mechanisms are slow as compared with typical time scales in
our system. 
Other single-particle states of the QD, above the exciton energy, can
be neglected because they are not strongly coupled to the cavity mode in the
case of resonance or quasi-resonance between the exciton and the confined photon. 
Under these conditions, the problem reduces to that of a two-level artificial atom
embedded in the optical cavity. We are going to show that, in spite of this
simplification, the system presents a very rich variety of physical situations.

The problem of a two-level system coupled to a single cavity mode, under different
conditions and approximations, has received a lot
of attention, mainly in the field of quantum optics 
\cite{cohen,walls,scully,yamamoto,kozlovskii,benson2}.
Of particular interest is the study of the possible sub-poissonian radiation when
cavity losses and pumping dominate onto the spontaneous emission of leaky 
modes. The aim of our work is an exhaustive analysis of the
different regimes of parameters to determine the role played by the different
physical mechanisms in the quantum properties of both the internal state of the
system and the light emitted in the steady regime under 
continuous incoherent pumping.

The paper is organized as follows: In section II we describe the model 
Hamiltonian and master equation that allows to
calculate the evolution of the populations and coherences of the energy levels.
By means of the quantum regression theorem, we can use our master equation to 
calculate cavity photon statistics and the spectrum of the emitted light.
In section III we present
the results for the time evolution of the magnitudes of interest, in different
ranges of parameters characterizing different physical situations. The
spectrum of the light emitted by the system is presented and discussed in
section IV. Section V is devoted to a summary of the work.

\section{Theoretical Framework}
We consider a single QD inside a semiconductor microcavity that is 
continuously and incoherently pumped. 
Our system is initially in its ground state and evolves until it reaches 
the steady-state. In order to describe the time evolution of the QD-cavity system 
we use a master equation that includes the strong exciton-photon coupling, 
the non-resonant pump, and the interaction with the environment that is 
responsible for dissipation.

\subsection{The Hamiltonian}
The physics of a QD strongly coupled with a single cavity mode is described by 
the following Hamiltonian: 
\begin{eqnarray}
H= H_{S} + H_{RS} + H_{R}
\label{hamilt}
\end{eqnarray}
$H_S$, $H_R$ are the Hamiltonian for the QD-cavity system, and the environment 
(reservoirs), respectively. The term $H_{RS}$ describes the interaction 
between the QD-cavity system and the reservoirs.
$H_S$ is given by the usual Jaynes-Cummings Hamiltonian \cite{walls,yamamoto}
that describes the 
interaction of a two-level system with a single mode of the electromagnetic field:
\begin{eqnarray}
H_S &=& H_0 + H_{X-C}
\nonumber \\
H_0 &=& \omega_X |X \rangle \langle X| + (\omega_X - \Delta ) a^{\dagger} a 
\nonumber \\ 
H_{X-C} &=& g \left( \sigma a^{\dagger} + a \sigma^{\dagger} \right)
\label{Jaynes-Cummings}
\end{eqnarray}
We have introduced ladder operators 
$\sigma^{\dagger} = \mid X \rangle \langle G \mid$, $\sigma = 
\mid G \rangle \langle X \mid$ connecting the ground $| G \rangle$ and excited 
(exciton) $| X \rangle$ states of the QD with energies zero and $\omega_X$ 
respectively (we take $\hbar =1$). $a^{\dagger}$ creates a cavity 
photon with energy $(\omega_X - \Delta)$. The coupling term $H_{X-C}$ in Eq. 
(\ref{Jaynes-Cummings}) describes the exciton-photon coupling in the Rotating Wave 
Approximation \cite{cohen,walls,scully}. 

$H_{RS}$ contains the coupling to external reservoirs including the 
following three process: 
\begin{itemize}
\item[(i)] The continuous and 
incoherent pumping of the QD by annihilating ($c_{R'}$) an electron-hole pair
from an external reservoir $R'$ (representing either electrical injection or
the capture of excitons optically created at frequencies larger than the typical
ones of our system) and creation ($d ^\dagger _{R''}$) a phonon emitted to a
reservoir $R''$ in order to take care of energy conservation
\begin{eqnarray}
\sum _{R',R''} \mu _{R',R''} [d^\dagger _{R''} c_{R'} \sigma ^\dagger
+ \sigma c^\dagger _{R'} d _{R''}] ,
\label{hpumping}
\end{eqnarray}
\item[(ii)]
The direct coupling of the QD exciton to the leaky modes, that is, to the 
photonic modes, with energy different than the cavity mode, that have a 
residual density of states inside the microcavity. This process is responsible 
for the dissipation of the excitonic degrees of freedom by  
the spontaneous emission to an external reservoir $R$ of photons 
created by $b ^\dagger _R$
\begin{eqnarray}
\sum _R \lambda _{lR} [\sigma b_R^\dagger +
b_R \sigma ^\dagger ] ,
\label{hleakyemiss}
\end{eqnarray}    
\item[(iii)] The escape of the cavity mode out of the microcavity due to 
the incomplete reflectance of the mirrors. The cavity mode is thus coupled to the 
continuum of photonic modes out of the microcavity. This process produces 
the direct dissipation of the cavity mode
\begin{eqnarray}
H_{CE}=\sum _R \lambda _R (ab_R^\dagger +b_Ra^\dagger ).
\label{hcavemiss}
\end{eqnarray}
\end{itemize}
The last term in Eq. (\ref{hamilt}),
$H_{R}$, describes the external reservoirs of harmonic oscillators (photons, phonons, 
electron-hole pairs,...) not being necessary to detail them explicitly. 
The three terms  $H_{RS}$ have coupling constants $\mu _{R',R''}$, $\lambda _R$ and
$\lambda _{lR}$ which depend on the particular mode ($R$, $R'$ or $R''$) of each
external reservoir. The operators $a$, $b_R$, $c_{R'}$ and $d_{R''}$ have bosonic
commutation rules as it corresponds to harmonic oscillators.
Since we are interested in the strong coupling regime, the first three terms,
$H_{0}$ and $H_{X-C}$ are treated exactly. That means that one could
work in the framework usually known as the "dressed
atom picture". However, in order to clarify the different pumping and losses
mechanisms, it is preferable to work in basis $\{ \mid Gn \rangle ;
\mid Xn \rangle \} $ in terms of the number of cavity photons $n$ and
bare ground $G$ and exciton $X$ states of the QD.

\subsection{Master equation}
We have made the whole algebra for a general case in which reservoirs are at 
finite temperature. However, the main physics already occurs for zero temperature,
which is the case we present hereafter in this paper.

We define the reduced density matrix, $\rho$, for the exciton-photon system 
by tracing out the reservoir degrees of freedom in the total density 
matrix $\rho _T$:
\begin{equation}
\rho = Tr_R(\rho_T) .
\label{reduced}
\end{equation}
In the interaction picture with respect to $H_{X-C} + H_{RS}$ in 
Eqs. (\ref{hamilt}) and (\ref{Jaynes-Cummings}), $\rho$ satisfies the 
master equation\cite{walls,scully}:
\begin{eqnarray}
\frac{d}{dt} \rho &=& \frac{i}{\hbar} \left[ \rho, H_S \right] + 
\nonumber \\
& & \frac{\kappa}{2} \left( 2 a \rho a^{\dagger} - a^{\dagger} a \rho - 
\rho a^{\dagger} a  \right) +
\nonumber \\
& & \frac{\gamma}{2} \left( 2 \sigma \rho \sigma^{\dagger} - \sigma^{\dagger} 
\sigma \rho - \rho \sigma^{\dagger} \sigma \right) + 
\nonumber \\
& & \frac{P}{2} \left( 2 \sigma^{\dagger} \rho \sigma - \sigma \sigma^{\dagger} 
\rho - \rho \sigma \sigma^{\dagger} \right) .
\label{masterequation}
\end{eqnarray}
The master equation is obtained under the usual Born-Markov approximation 
for the interaction $H_{RS}$ between the QD-cavity system and the reservoirs, but 
the strong exciton-cavity photon coupling, $H_{X-C}$, is described exactly. 
$\kappa$ is the decay of the cavity photon by escaping through the 
microcavity mirrors, $\gamma$ is the decay of the QD exciton by the 
spontaneous emission into leaky modes, and $P$ is the rate of continuous 
incoherent pumping of the QD exciton. By means of Eq. (\ref{masterequation}) 
one can get a set of differential equations that describe the evolution of 
the populations and coherences of the cavity-QD system. In 
the basis $\{ |Gn \rangle ; | Xn \rangle \}$ of product 
states between QD states and Fock states of the cavity mode, 
the matrix elements of the reduced density matrix are:
\begin{equation}
\rho_{in,jm} = \langle i n | \rho |j m \rangle ,
\label{matrixelements}
\end{equation}
with $i, j$ being either $G$ or $X$.
The diagonal matrix elements $\rho_{Gn, Gn}$, $\rho_{Xn, Xn}$ are the 
populations of the QD-photon levels, while the 
non-diagonal terms  $\rho_{G n,X n-1}$, $\rho_{X n-1, G n}$, describe the 
coherences between these levels. 
By taking the matrix elements in Eq. ($\ref{masterequation}$) we get, for $T=0$,
the following set of linear differential equations:
\begin{figure}[H]
\includegraphics [clip,height=8cm,width=10.cm]{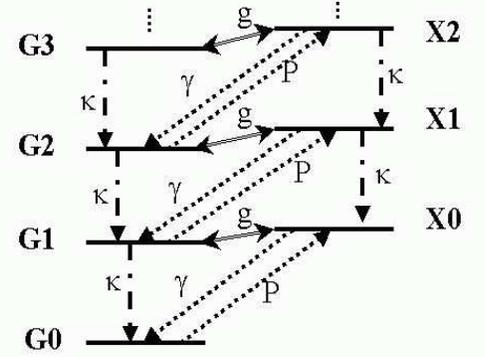}
\label{scheme}
\caption{Ladder of levels (continuous lines) for a two-state QD coupled to a single
optical mode of a microcavity. States labeled as $Gn$ with $n=0,1,2,...$ correspond
to having the QD in its GS coexisting with $n$ cavity photons. The same for
states $Xn$ with the QD in its excited state $X$. The double, dashed and dash-dotted
lines depict the coupling, pumping and emission processes with rates $g$, $P$,
$\gamma  $ and $\kappa$ explained in the text.}
\end{figure}
\begin{widetext}
\begin{eqnarray}
& \partial _{t} \rho _{G n, G n} & = i g \sqrt{n} \left( \rho _{G n, X n-1}-
\rho _{X n-1, G n} \right) \nonumber \\
& & + \gamma \rho _{X n, X n}-\kappa \left[ n \rho _{G n, G n}- (n+1) 
\rho _{G n+1, G n+1}
\right] -P \rho _{Gn, Gn}
\label{rho1}
\end{eqnarray} \begin{eqnarray}
& \partial _{t} \rho _{X n, X n} &= i g \sqrt{n+1} \left( \rho _{X n,G n+1}-
\rho _{G n+1, X n} \right) \nonumber \\
& & - \gamma \rho _{X n, X n} - \kappa \left[ n \rho _{X n, X n}- (n+1) 
\rho _{X n+1, X n+1}
\right] + P \rho _{G n, G n}
\label{rho2}
\end{eqnarray} \begin{eqnarray}
&\partial _{t} \rho _{G n, X n-1} & = i \left[ g \sqrt{n} \left( \rho _{G n, G n} -
\rho _{X n-1, X n-1}\right) + \Delta \rho _{G n, X n-1} \right] \nonumber \\
& & -\left[ \left( \gamma +\kappa (2n-1)+ P \right) /2 \right] \rho _{G n, X n-1}
+\kappa \sqrt {n(n+1)} \rho _{G n+1, X n} ,
\label{rho3}
\end{eqnarray}
plus the equation hermitian conjugate of (\ref{rho3}).
In the absence of dissipation, the matrix elements $\rho_{G n, G n}$, 
$\rho_{X n-1, X n-1}$, $\rho_{G n,X n-1}$, and $\rho_{X n-1, G n}$, for a 
given photon number $n$, satisfy a closed set of four differential equations. 
However, the pumping and emission with rates $P$, $\kappa$ and $\gamma$, 
couple the terms with different 
photon occupation number $n$, so that an infinite set of equations has to 
be solved, as depicted schematically in Fig. 1. For the numerical integration, 
the set of equations can be truncated at a given value that,
in all the cases considered below, it is enough to take equal to 100.

As initial conditions we take $\rho_{G0,G0}=1$ and all the
other elements of the density matrix equal to zero.
In other words, we start with the system in its ground state and the
pumping, and subsequently the losses, produces the dynamical evolution
of the whole system. The steady-state of the system can be studied by integrating 
the set of equations (\ref{rho1}), (\ref{rho2}) and (\ref{rho3}) for long times.

\subsection{First and second order coherence functions}
From the master equation, one can compute the dynamics of the expectation values
of any operator. Moreover, two-time correlation functions are also of physical interest.
In particular we want to compute the first and second order coherence 
functions\cite{scully,yamamoto}
\begin{eqnarray}
g^{(1)}(\vec{r},t,\tau )= \frac{\langle E^{(-)}(\vec{r},t)E^{(+)}(\vec{r},t+\tau)
\rangle}{\sqrt{\langle E^{(-)}(\vec{r},t)E^{(+)}(\vec{r},t) \rangle
\langle E^{(-)}(\vec{r},t+\tau)E^{(+)}(\vec{r},t+\tau) \rangle}}
\end{eqnarray}
\begin{eqnarray}
g^{(2)}(\vec{r},t,\tau )= \frac{\langle E^{(-)}(\vec{r},t)E^{(-)}(\vec{r},t+\tau)
E^{(+)}(\vec{r},t+\tau) E^{(+)}(\vec{r},t)\rangle}{\langle E^{(-)}(\vec{r},t)
E^{(+)}(\vec{r},t) \rangle
\langle E^{(-)}(\vec{r},t+\tau)E^{(+)}(\vec{r},t+\tau) \rangle}
\label{g2}
\end{eqnarray}
\end{widetext}
where $E^{(+)}$ and $E^{(-)}$ are the positive and negative frequency parts of
the amplitude of the electromagnetic field. In the steady-state
limit we are interested in, both $g^{(1)}$ and $g^{(2)}$ do not depend on 
the absolute time $t$.

The experimental situation is such that photonic modes escaping from the cavity
are emitted in well defined direction, for instance along the axis of the
micropillars in the direction in which the mirrors defining the cavity
have some transparency\cite{gerard,benson,solomon,moreau,santori,pelton}.
On the contrary, emission of leaky modes takes place in any direction,
particularly through lateral surfaces of the micropillar. This means that
simply by changing the spatial distribution of detectors, one can measure the
emission from cavity modes or the leaky modes.

By Fourier transforming the first order correlation function
\begin{eqnarray}
G^{(1)}(t,\tau)=\langle E^{(-)}(\vec{r},t)E^{(+)}(\vec{r},t+\tau) \rangle ,
\label{G1}
\end{eqnarray}
one can obtain the the power spectrum of the emitted light.
\begin{eqnarray}
S(\vec{r},\nu)= \frac{1}{\pi} \Re \int _0 ^\infty d \tau e^{i \nu \tau}
G^{(1)}(t,\tau) .
\label{intensity1}
\end{eqnarray}
The first order correlation function of the external field can be obtained
from the time dependence of the operators describing intrinsic properties
of the system: 
\begin{eqnarray}
G^{(1)}_{C}(t, \tau) & = &  \langle a^{\dagger }(t+\tau) a(t) \rangle 
\nonumber \\
G^{(1)}_{X}(t, \tau) & = &  \langle \sigma^{\dagger}(t+\tau) \sigma(t) \rangle.
\label{intensity2}
\end{eqnarray}
$G^{(1)}_{C}(t, \tau)$ is the correlation function for the cavity mode, 
responsible for the stimulated emission part of the spectrum (i.e.
the light coming from the confined photon)  
by means of $G^{(1)}(t,\tau) \propto \kappa G^{(1)}_{C}(t, \tau) $.
In the case of pillar microcavities, it would correspond to the light emitted
in the vertical direction. On the other hand $G^{(1)}_{X}(t, \tau)$ describes
the spontaneous part of the spectrum (i.e. light directly coupled to the QD exciton)
by means of $G^{(1)}(t,\tau) \propto \gamma G^{(1)}_{X}(t, \tau) $. This gives
the light emitted through the leaky modes that can be measured in the lateral 
direction of a micropillar.

In order to calculate these two-time correlation functions, we make use of the
quantum regression theorem \cite{walls,scully,yamamoto}. 
First of all, we define the following operators:
\begin{eqnarray}
a^\dagger _{G n} (\tau ) &=&  \mid G n+1 \rangle \langle G n \mid 
e^{i (\omega - \Delta) \tau } \nonumber \\
a^\dagger _{X n} (\tau ) &=&  \mid X n+1 \rangle \langle X n \mid 
e^{i (\omega - \Delta) \tau } \nonumber \\
\sigma ^\dagger_n (\tau ) &=&  \mid X n \rangle \langle G n \mid 
e^{i \omega  \tau } \nonumber \\
\varsigma _{n}(\tau ) &=& \mid G n+1 \rangle \langle X n-1\mid
e^{i (\omega -2\Delta) \tau }.
\label{operator}
\end{eqnarray}
In the interaction picture, time evolution with respect to $H_{X-C} + H_{RS}$ 
appears explicitly. The two-time correlation function of the 
cavity mode can be expressed in terms of these operators:
\begin{eqnarray}
& G^{(1)}_{C}(t, \tau) & = 
\sum_n \sqrt{n+1} \times 
\\
& & \left( \langle a^{\dagger}_{G n}(t+\tau) a(t) \rangle +
\langle a^{\dagger}_{X n}(t+\tau) a(t) \rangle \right) . \nonumber
\label{2timecav}
\end{eqnarray} 
The exciton correlation function can be written as:
\begin{equation}
G^{(1)}_{X}(t, \tau)=
\sum_n \langle 
\sigma^{\dagger}_n(t+\tau) \sigma(t) \rangle
\label{2timeexc1}
\end{equation}
For the calculation of the spectrum of the emitted light it is, thus, necessary 
to evaluate the functions 
$\langle a^{\dagger}_{G n}(t+\tau) a(t) \rangle$, $\langle a^{\dagger}_{X n}(t+\tau) 
a(t) \rangle$, and 
$\langle \sigma^{\dagger}_n(t+\tau) \sigma(t) \rangle$. The quantum regression 
theorem states that given a set of operators $O_j$, whose averages satisfy 
a closed set of linear differential equations:
\begin{equation}
\frac{d}{d \tau} \langle O_j (t+\tau) \rangle = \sum_k L_{j,k} \langle O_k 
(t+\tau ) \rangle , 
\label{qrth1}
\end{equation}
then the two-time averages of $O_j$ with any other operator $O$, also satisfy 
the same differential equation:
\begin{equation}
\frac{d}{d \tau} \langle O_j(t+\tau) O(t) \rangle = \sum_k L_{j,k} \langle O_k 
(t+\tau) O(t) \rangle
\end{equation}
In order to get the time evolution of two-time averages, we start with the 
dynamics of the operators (\ref{operator}).
Their averages satisfy the 
set of linear differential equations that allows us to find the evolution of 
the corresponding two-time averages in (\ref{2timecav}, \ref{2timeexc1}):
\begin{widetext}
\begin{eqnarray}
\partial _\tau \langle a^\dagger _{G n} (\tau ) \rangle &=&
\left( \partial _\tau \rho _{G n,G n+1} \right) e^{i (\omega - \Delta) \tau }+
i (\omega -\Delta ) \langle
a^\dagger _{G n} (\tau ) \rangle \nonumber \\
\partial _\tau \langle \sigma ^\dagger_n (\tau ) \rangle &=&
\left( \partial _\tau \rho _{G n, X n} \right) e^{i \omega\tau }+
i \omega \langle \sigma ^\dagger_n (\tau ) \rangle  \nonumber \\
\partial _\tau \langle a^\dagger _{X n-1} (\tau ) \rangle &=&
\left( \partial _\tau \rho _{X n-1, X n} \right) e^{i(\omega - \Delta) \tau }+
i (\omega -\Delta ) \langle a^\dagger _{X n-1} (\tau ) \rangle \nonumber \\
\partial _\tau \langle \varsigma _{n}(\tau ) \rangle &=&
\left( \partial _\tau \rho _{X n-1,G n+1} \right) e^{i (\omega - 2\Delta) \tau }+
i (\omega -2\Delta ) \langle \varsigma _{n}(\tau ) \rangle .
\end{eqnarray}
Although the operator $\varsigma _{n}$ is not needed in the evaluation of the 
required two-time averages (\ref{2timecav}) and (\ref{2timeexc1}),
we have to add the correlation functions that include such operator
$\varsigma_n$ in order to get a closed set of equations. 
Using the master equation and eliminating the elements of the density matrix, 
we arrive at the desired time-evolution for the averages (\ref{operator}): 
\begin{eqnarray}
& \partial _\tau \langle a^\dagger _{G n} (\tau ) \rangle =&
\langle a^\dagger _{G n} (\tau ) \rangle \left[ 
i(\omega_X-\Delta) - \frac{\kappa}{2}(2n+1)-P \right]
\nonumber\\
& &+\langle a^\dagger _{G n+1} (\tau ) \rangle \kappa\sqrt{(n+1)(n+2)}
+\langle \sigma ^\dagger_n (\tau ) \rangle ig\sqrt{n+1} \nonumber\\
& &+\langle a^\dagger _{X n-1} (\tau ) \rangle \gamma -
\langle \varsigma _{n}(\tau ) \rangle ig\sqrt{n} 
\nonumber \\
& &
\nonumber \\
& \partial _\tau \langle \sigma ^\dagger_n (\tau ) \rangle = &
\langle a^\dagger _{G n} (\tau ) \rangle ig\sqrt{n+1} +
\langle \sigma ^\dagger_n (\tau ) \rangle \left[
i\omega_X -\frac{(\gamma+P)}{2}-\kappa n \right] \nonumber \\
& & +\langle \sigma ^\dagger_{n+1} (\tau ) \rangle \kappa (n+1) -
\langle a^\dagger _{X n-1} (\tau ) \rangle ig\sqrt{n}
\nonumber \\
& &
\nonumber \\
& \partial _\tau \langle a^\dagger _{(X n-1)} (\tau ) \rangle = &
\langle a^\dagger _{G n} (\tau ) \rangle P -
\langle \sigma ^\dagger_n (\tau ) \rangle ig\sqrt{n} \nonumber \\
& &+\langle a^\dagger _{X n-1} (\tau ) \rangle \left[ 
i(\omega_X-\Delta)-\gamma-\frac{\kappa}{2} (2n-1) \right]
\nonumber\\
& &+\langle a^\dagger _{X n} (\tau ) \rangle \kappa \sqrt{n(n+1)} +
\langle \varsigma _{n}(\tau ) \rangle ig\sqrt{n+1} 
\nonumber \\
& &
\nonumber \\
& \partial _\tau \langle \varsigma _{n}(\tau ) \rangle = &
-\langle a^\dagger _{(G n)} (\tau ) \rangle ig\sqrt{n} +
\langle a^\dagger _{(X n-1)} (\tau ) \rangle ig\sqrt{n+1} \nonumber \\
& & +\langle \varsigma _{n}(\tau ) \rangle \left[ 
i(\omega_X-\Delta) -\frac{(\gamma+P)}{2}-\kappa n \right] \nonumber\\
& & +\langle \varsigma _{n+1}(\tau ) \rangle \kappa \sqrt{n(n+2)} .
\label{eq-operators}
\end{eqnarray}
From Eq. (\ref{eq-operators}), and the quantum regression theorem, 
it is straightforward to obtain two separate closed sets of differential 
equations for two-times functions: one for the set of functions 
\begin{equation}
\{ \langle a^\dagger _{G n}(\tau +t) a(t) \rangle, 
\langle a^\dagger _{X n-1}(\tau +t) a(t) \rangle, 
\langle \sigma^{\dagger}_{n}(\tau +t) a(t) \rangle, 
\langle \varsigma_{n}(\tau +t) 
a(t) \rangle \}, 
\label{set1}
\end{equation}
needed for the calculation of $G^{(1)}_{C}(t, \tau)$, and other 
for the set of functions 
\begin{equation}
\{ \langle a^\dagger _{G n}(\tau +t) \sigma(t) \rangle, 
\langle a^\dagger _{X n-1}(\tau +t) \sigma(t) \rangle, 
\langle \sigma^{\dagger}_{n}(\tau +t) \sigma(t) \rangle, 
\langle \varsigma_{n}(\tau +t) 
\sigma(t) \rangle \},
\label{set2}
\end{equation}
needed for the calculation of $G^{(1)}_X(t, \tau)$.

\end{widetext}
An important point is the initial conditions to solve these systems of equations. 
Such conditions are obtained by solving, up to
the stationary limit, the master equations (\ref{rho1}), (\ref{rho2})
and (\ref{rho3}) for the density matrix. From this information,
one knows the initial conditions. For the functions in (\ref{set1}):
\begin{eqnarray}
\langle a^{\dagger}_{G n}(t) a(t) \rangle &=& \sqrt{n+1} \rho_{G n+1,G n+1}
\nonumber \\
\langle a^{\dagger}_{X n-1}(t) a(t) \rangle &=& \sqrt{n} \rho_{X n, X n} 
\nonumber \\
\langle \sigma^{\dagger}_n (t)  a(t) \rangle &=& \sqrt{n+1} \rho_{G n+1, X n}
e^{-i \Delta t}
\nonumber \\
\langle \varsigma_n(t) a(t) \rangle &=& \sqrt{n} \rho_{X n, G n+1}
e^{i \Delta t}.
\label{ic1}
\end{eqnarray}
For the functions in (\ref{set2}):
\begin{eqnarray}
\langle a^{\dagger}_{G n}(t) \sigma(t) \rangle &=& \rho_{X n, G n+1}
e^{-i \Delta t}
\nonumber \\
\langle a^{\dagger}_{X n-1}(t) \sigma(t) \rangle &=& 0
\nonumber \\
\langle \sigma^{\dagger}_n \sigma(t) \rangle &=& \rho_{X n, X n}
\nonumber \\
\langle \varsigma _{n} \sigma(t) \rangle &=& 0
\label{ic2}
\end{eqnarray}

The second order coherence function is, a priori, more complicated to calculate.
Averages of products of four operators at
two different times have to be performed. This task becomes much simpler 
for the case of zero time delay.
$g^{(2)}(t,\tau=0)$ is a one-time operator, which simplifies the calculation.
In spite of loosing information, $g^{(2)}(t,\tau=0)$ is a very interesting magnitude
because it can be used as indicator of the possible coherence of the
state of the system\cite{cohen,walls,scully,yamamoto}.
In addition, if we concentrate in the properties of cavity modes (i.e., neglecting
the emission of leaky modes), the second order coherence function
acquires a very simple form in the stationary limit:
\begin{eqnarray}
& g^{(2)} (t,\tau=0) & = \frac {\langle a^\dagger a^\dagger a a \rangle}
{ \langle a^\dagger a \rangle ^2} \\
& & = \frac{\sum_n n(n-1) [\rho_{X n,X n} + \rho_{G n,G n}]}
{[\sum_n n[\rho_{X n, X n} +\rho_{G n, G n}]]^2} . \nonumber
\label{g22}
\end{eqnarray}
$g^{(2)}$ takes different values depending on the statistics of the photon state:
$g^{(2)} = 2$ for chaotic states, $g^{(2)}=1$ for states having a Poisson 
distribution in $n$, and 
$g^{(2)}<1$ for non classical systems having a sub-Poisson distribution
\cite{walls,scully,yamamoto}.

\section{Results}
\subsection{Election of the range of parameters of interest}

Since our model has several parameters, we need some simple picture in order to get
insight on the interesting regime of parameters.
Although our system is similar to a spin coupled to an harmonic oscillator,
we can understand the effect of dissipation by considering two-coupled 
(by $g$) oscillators.
One of them has an eigenfrequency $\omega _X - \Delta $ being damped with a
rate $\kappa $ while the other, with an eigenfrequency
$\omega _X $, is damped with a rate $\gamma$ and is pumped incoherently
with a rate $P$. The incoherent pumping implies a dephasing mechanism which,
together with the damping, gives an average dephasing $(\gamma+P)/2$. When the
damping of the first oscillator $\kappa $ is much higher than the dephasing
$(\gamma+P)/2$ of the second oscillator, one could expect that the frequency of
this second oscillator $\omega _X $ will survive longer than $\omega _X-\Delta $
and it will dominate the spectrum. In the opposite limit, $\kappa \ll (\gamma +P)/2$,
the frequency $\omega _X - \Delta $ will dominate the spectrum. This
behavior is the one shown by our spin-oscillator system as can be checked by means
of the Haken's model on adiabatic elimination \cite{walls,scully,gardiner}
in the equations of motion of the field operators.
From this simple picture we can motivate the existence of three
different regimes $\kappa \gg (\gamma +P)/2$, $\kappa \ll (\gamma +P)/2$ and
some intermediate regime. For this purpose, we fix $g=1$ as the energy scale, take
a fixed small value $\gamma =0.1$ reflecting the fact that leaky modes emission
is the less efficient mechanism in practical situations, and define the three
regimes by changing $\kappa$ and $P$.

Apart from those parameters, our model requires to set the frequency of the two
oscillators. In practical cases, the coupling between the oscillators is in the
order of meV while the excitation energy $\omega_X$ is in the order of eV.
Therefore, since our energy scale is $g=1 (meV)$, all the calculations we present here
have been performed by taking $\omega_X =1000$. Finally the detuning $\Delta$ is
a very important parameter. Since typical detunings are in the order of a few meV,
we present here results for two different cases: perfect resonance, $\Delta =0$
and quasi-resonance $\Delta =5$.

\subsection{Dynamics of the density matrix: occupations and coherences}
In this subsection we show the time evolution of the occupations and coherences
(diagonal and off-diagonal respectively) $\rho_{i n,j m}$ described in Eqs.
(\ref{rho1}),(\ref{rho2}) and (\ref{rho3}).
From these equations, it is clear that if losses and pumping are not included,
the systems shows the usual Rabi oscillations in each subspace with a given 
number of photons.  The amplitude of these oscillations increases with the 
coupling, $g$, and decreases with the detuning $\Delta $.

When all the losses and the pumping are included, the situation changes
drastically.
The initial occupation of $\mid G0\rangle $ evolves with time
up to a state of equilibrium in which occupation is
redistributed among a large number of levels with a finite number of photons.
In this final steady situation, the sum of all the losses equilibrates the pumping.
\begin{figure}
\psfig{figure=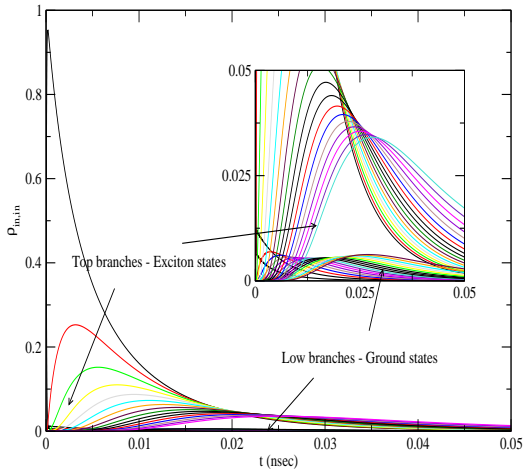,height=8.0cm,width=8.0cm,angle=-90}
\label{diagonal1}
\caption{Density matrix diagonal elements $\rho _{in,in}$
for both the exciton ($i\equiv X$)
and the ground ($i\equiv G$) states, from $n=0$ up to $n=35$ cavity photons.
$g=1$, $\Delta=5$, $\gamma=0.1$, $\kappa=0.1$ and $P=15$. The inset shows
a zoom of the graphic to clarify the formation of two bundle of branches
associated with the exciton and the ground states of the QD respectively.}
\end{figure}

Figs. 2 and 3 show the time evolution of occupations $\rho _{i n, i n}$ for
both the exciton ($i\equiv X$) and the ground ($i\equiv G$) states.
The results in Figs. 2 and 3 are typical of the two extreme regimes described 
above. When pumping rate $P$ dominates (Fig. 2), states with a large
number of photons $n$ become occupied and two bundles of branches can be distinguished:
upper branches corresponding to $\mid X n \rangle $ states and lower branches
corresponding to $\mid G n \rangle $ states. In any case, the differences between
the occupations of states in the two bundles, is not very large. When rates
for losses increase and become comparable to the pumping rate, just a few photons can
be stored in the cavity. In equilibrium, the highest occupation is that of the
$\mid G 0 \rangle $ state and only states with $n < 3$ have appreciable,
although small, occupations.
It is quite evident that an interesting regime is missing in these results: the
one in which losses dominate onto the pumping, for instance the case $\kappa =5$
and $P=1$. We will discuss below other characteristics of this case but do not
include a figure here because the corresponding time dependence is featureless
showing a fast decay to zero of the occupations for all the states with $n > 0$.

Time evolution of coherences $\rho _{X0,G1}$ is shown in Fig. 4 for the two
limiting cases $P \gg \kappa$ and $P \ll \kappa$ for $\Delta =5$. Due to this 
finite detuning, Rabi-like oscillations appear in the coherences. 
When $\kappa $ dominates, the
coherences increase with time up to finite, but small, value. 
When pumping dominates, the mean number of photons inside the cavity 
is much higher than 1 as discussed below. Therefore, a coherence like the one 
in Fig. 4 between states with $n=0$ and $n=1$ decreases with time. In this case, 
$P \gg \kappa$, we have computed coherences $\rho _{Xn, Gn+1}$, for values of 
$n$ around the mean number of photons inside the cavity (see below), obtaining 
that they also go to a small constant (apart from the Rabi oscillations) for 
large times.
\begin{figure}
\psfig{figure=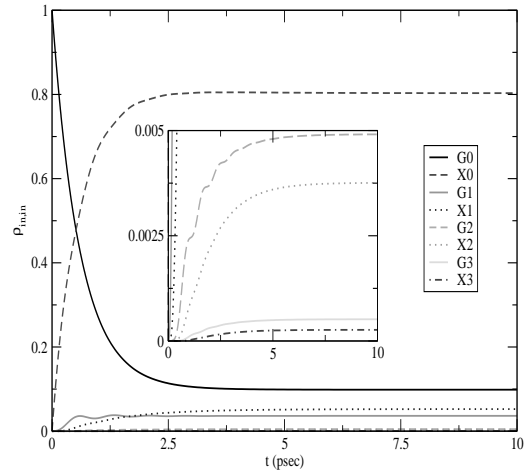,height=8.0cm,width=8.0cm,angle=-90}
\label{diagonal2}
\caption{Density matrix diagonal elements $\rho _{in,in}$
for both the exciton ($i\equiv X$)
and the ground ($i\equiv G$) states, from $n=0$ up to $n=3$ cavity photons.
$g=1$, $\Delta=5$, $\gamma=0.1$, $\kappa=0.5$ and $P=1$. The inset shows a
zoom of the vertical axis of the graphic.}
\end{figure}
\begin{figure}
\psfig{figure=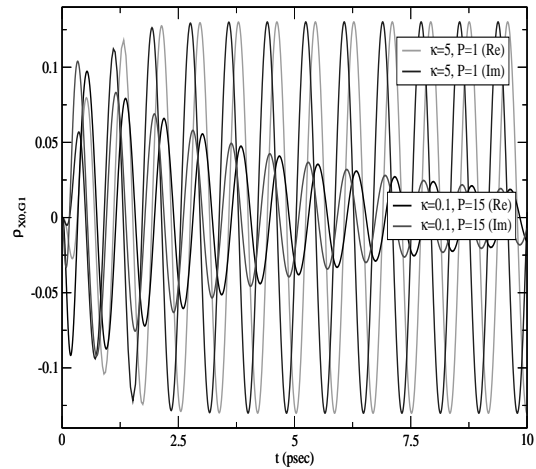,height=8.0cm,width=8.0cm,angle=-90}
\label{nodiagonal2}
\caption{Density matrix off diagonal element $\rho _{X0,G1}$ with detunning
$\Delta=5$ for $g=1$, $\gamma=0.1,$ and two different values of $\kappa$ and $P$.}
\end{figure}

All the results included in Figs. 2, 3 and 4 correspond to a finite detuning
$\Delta =5$. In the case of perfect resonance $\Delta =0$, the main effect is
the lack of oscillations as shown, as an example, in Fig. 5 for the coherences
$\rho _{X0,G1}$.

The fact, shown above, that $P$ competes with $\kappa$, is not completely 
general. This fact can be seen in equations (\ref{rho1}), (\ref{rho2}) and
(\ref{rho3}) for the time evolution of the density matrix elements.
For the off-diagonal terms, $P$ has the same sign than the losses $\kappa $ and 
$\gamma $, while for the diagonal terms, these signs are different.  
The pumping produces decoherence in the off-diagonal terms as 
the emission of photons does because in our model the pumping is
incoherent. At the same time, $P$ favors higher occupations (increase of the 
diagonal elements).
Therefore it is clear that, for a high pumping, decays of the off-diagonal 
terms (loss of coherence) prevail on the occupation until
the coherence is completely quenched and later inverted, preventing the evolution
towards the occupation of states with a high number of photons. This is an
indication of the important role that off-diagonal elements play in the time
evolution. It also explains that, if the emission rate $\kappa$ is above a
critical value, it dominates on any pumping effect.
\begin{figure}
\psfig{figure=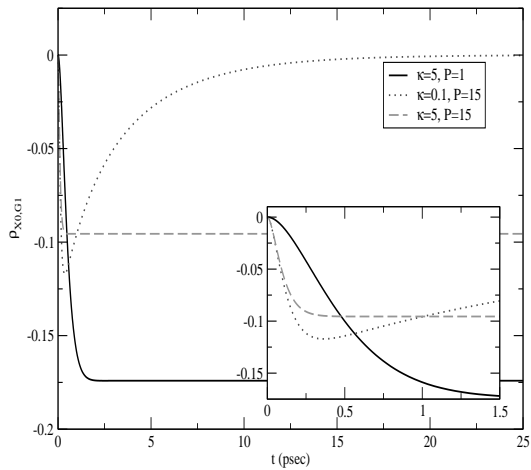,height=8.0cm,width=8.0cm,angle=-90}
\label{nodiagonal1}
\caption{Imaginary part of the density matrix off diagonal element $\rho _{X0,G1}$
at perfect resonance ($\Delta=0$)
for $g=1$, $\gamma=0.1,$ and different values of $\kappa$ and $P$.
All the real parts of the same matrix elements are zero. The inset shows a
zoom of the graphic for small values of $t$.}
\end{figure}

\subsection{Number of cavity photons}

A very important result to be drawn from the time evolution discussed above is the
fact that the system is able to store a significant number of cavity photons.
Let us analyze the mean number of cavity photons
as a first step to study the quantum properties of these photons by looking
to its distribution. The mean number of photons in the cavity is
\begin{eqnarray}
\overline{N_{ph}}=\sum_n n(\rho_{X n,X n} + \rho_{G n, G n}) .
\label{numtph}
\end{eqnarray}
\begin{figure}
\includegraphics [clip,height=8cm,width=10.cm]{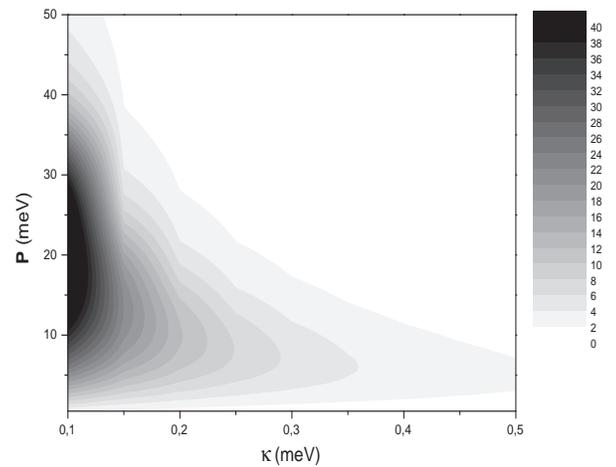}
\label{3DNph1}
\caption{$\overline{N_{ph}}$, in grey scale, as a function of $\kappa$ and $P$
for perfect resonance $\Delta=0$ with $g=1$ and $\gamma=0.1$.}
\end{figure}

\begin{figure}[H]
\includegraphics [clip,height=8cm,width=10.cm]{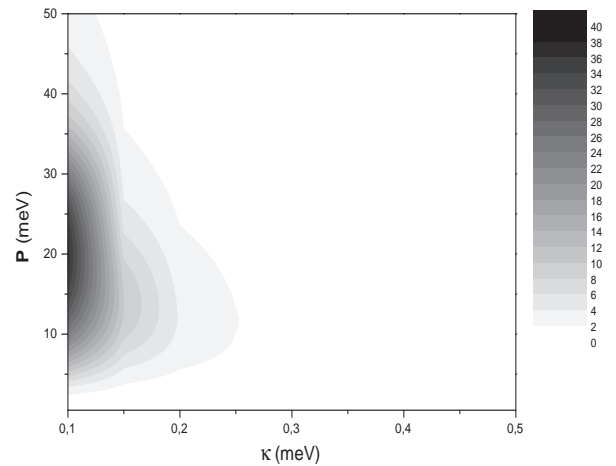}
\label{3DNph2}
\caption{$\overline{N_{ph}}$, in grey scale, as a function of $\kappa$ and $P$
for a detuning $\Delta=5$ with $g=1$ and $\gamma=0.1$.}
\end{figure}
$\overline{N_{ph}}$ increases with time, in a scale of hundreds of
picoseconds, up to the stationary value that is the magnitude we are interested in.
Figs. 6 and 7 show a contour plot of the mean number of cavity photons for
$\Delta =0$ and $\Delta =5$ respectively. As in previous figures, we fix all the
parameters except $P$ and $\kappa$, which vary along the two axis.
In these figures a region with a high $\overline{N_{ph}}$ can be observed.
It corresponds to low emission and high pumping. The region is larger and the
number of cavity photons higher for the
resonant case, $\Delta=0$, than for the case with detuning $\Delta=5$.
A detuning makes more difficult the coupling between the states $\mid X n
\rangle$ and $\mid G n+1 \rangle$. As a consequence the mechanism of storing cavity
photons become less efficient.

A conclusion similar to the one drawn in the previous section is in order: when
pumping increases, the mean number of cavity modes also increases up to a certain
maximum value. Further increase of $P$ implies that the dephasing mechanism
related with this incoherent pumping provokes a decrease of $\overline{N_{ph}}$.

\subsection{Second order coherence function}

The possibility of storing a large number of cavity photons for a wide range of
parameters opens an interesting alternative beyond the simple value of
$\overline{N_{ph}}$: how is the distribution of cavity photons? If any interesting
distribution is involved, its character could be transfered to the light emitted
by the system.

In order to quantify if such distributions are closer to gaussian distributions 
(thermal or chaotic states), Poisson distributions or even non-classical 
sub-poissonian distributions, one can compute, using Eq.
(\ref{g22}), the second order coherence function $g^{(2)} (\tau =0)$ discussed above.
\begin{figure}
\includegraphics [clip,height=8cm,width=10.cm]{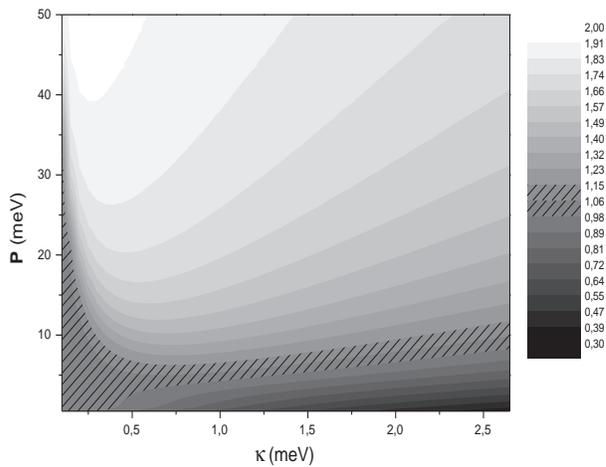}
\label{3Dg21}
\caption{Second order coherence function $g^{(2)}(\tau=0)$, in grey scale,
as function of $\kappa$ and $P$ for perfect resonance $\Delta=0$.
$g=1$ and $\gamma=0.1$. The dashed area shows the region supporting a
Poisson distribution of cavity photons in which $g^{(2)}(\tau=0)$ close to 1.}
\end{figure}

Figs. 8 and 9 show $g^{(2)} (0)$, for $\Delta=0$ and $\Delta=5$ respectively.
In these two figures we have dashed the region in which $g^{(2)} (0)$ is close to
1 to show the region supporting a poissonian distribution. This is the 
border between classically accessible region and the one having $g^{(2)} (0)< 1$ 
with non-classical states with sub-poissonian distribution.
In the case of perfect resonance (Fig. 8), for low $\kappa$, 
the poissonian distribution region is rather wide in terms of the pumping
$P$. When $\kappa$ increases, this region becomes much narrower in $P$.
In the case of detuning $\Delta=5$, the poissonian region becomes much smaller 
and even disappearing for intermediate values of $\kappa$.
\begin{figure}
\includegraphics [clip,height=8cm,width=10.cm]{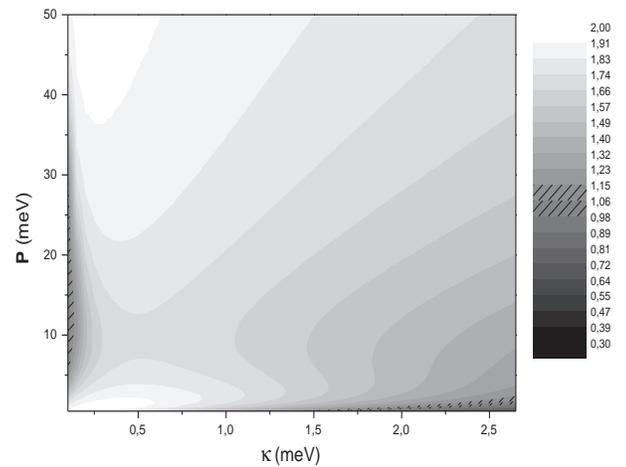}
\label{3Dg22}
\caption{Second order coherence function $g^{(2)}(\tau=0)$, in grey scale,
as function of $\kappa$ and $P$ for a detuning $\Delta=5$.
$g=1$ and $\gamma=0.1$. The dashed area shows the region supporting a
Poisson distribution of cavity photons in which $g^{(2)}(\tau=0)$ close to 1.}
\end{figure}

The interesting region of $g^{(2)} (0)<1$ appears for very low pumping in these 
two figures. Even though this occurs in a small region of parameters, it is very
interesting because it means that the system can form squeezed photon states with
sub-poissonian distribution. In this region the emitted light shows the antibunching
that is characteristic of non-classical light emitters. 

The rest of the diagram is supporting states with $g^{(2)} (0)$ increasing from 
1 to 2, i.e. states where second order coherence is reduced and
approaching a gaussian distribution ($g^{(2)} (0)=2$). A finite detuning  $\Delta $
enlarges this gaussian region as observed when comparing the two figures.

Combining and summarizing the results of Figs. 6, 7, 8 and 9, it must be
stressed that, for high quality cavities, i.e. low $\gamma$ and $\kappa$,
and intermediate value of the incoherent pumping $P$ is able to produce
a rather large number of cavity photons with Poisson distribution. When detuning 
increases this
effect remains, although with a reduction of the region of parameters supporting it.

\section{Spectrum of the emitted light}
In this section we will present our results for
the spectrum of the emitted light in the stationary limit ($t \rightarrow \infty $).
This is the property that can be most easily measured in experiments.
We will do it only in the case of finite detuning $\Delta =5$ in order to observe
the emission at two different frequencies. 
Following the steps discussed in the section devoted to our model, we have computed
$G^{(1)}_X (\tau)$ and $G^{(1)}_C (\tau)$ in the stationary limit ($t \rightarrow 
\infty $) for the three regimes already discussed $P \gg \kappa$,
$P \ll \kappa$ and $P$ comparable to $\kappa$. The envelopes (with fast 
oscillations inside) of these functions are shown in Figs.
10, 11 and 12 in which we show different contributions to the total emission.
In these three figures, the upper part shows the spontaneous emission of leaky
modes ($G^{(1)}_X (\tau)$), while the two lower parts show two different 
contributions to the
stimulated emission through cavity modes ($G^{(1)}_C (\tau)$).
Fig. 10 correspond to a case with a large $\overline{N_{ph}}$ inside the cavity. 
Therefore, when summing up in $n$ to get $G^{(1)}_X (\tau)$ and $G^{(1)}_C
(\tau)$, one obtains a time scale significantly larger than in the case of Figs. 11 
and 12, which correspond to a small $\overline{N_{ph}}$ inside the cavity.
\begin{figure}
\psfig{figure=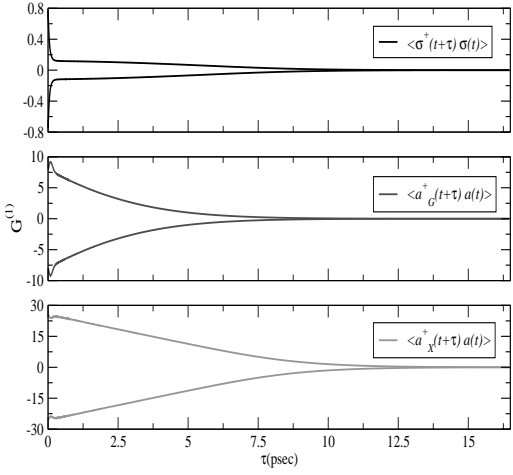,height=8.0cm,width=8.0cm,angle=-90}
\label{g1t1}
\caption{Envelope of the first order correlation functions $G_X^{(1)}(\tau)$
(upper panel) and $G_C^{(1)}(\tau)$ (two lower panels)
at the stationary limit ($t \rightarrow \infty $). $g=1$, $\Delta=5$,
$\gamma=0.1$, $\kappa=0.1$ and $P=15$.}
\end{figure}

The Fourier transform of Figs. 10, 11 and 12 gives the spectrum of the light
emission in the three cases discussed above. In order to numerically perform
such Fourier-transforms, we have used digital data filters (Parzen and Haning) 
to reduce numerical noise. The spectra corresponding to
$\omega _X= 1000$ are shown in Figs. 13, 14 and 15 . In these
three figures we have made use of the fact, discussed in the introduction, that
the two types of emissions can be separated:
the emission coming from the cavity modes takes place along the axis of the 
pillars, while the emission of leaky modes takes place in any spatial 
direction (mainly in the directions perpendicular to the axis of the pillar).
\begin{figure}
\psfig{figure=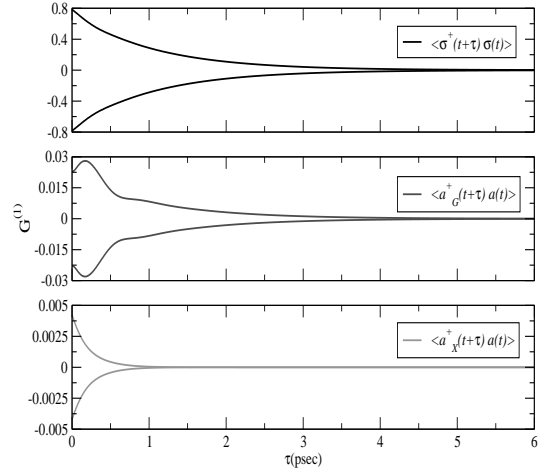,height=8.0cm,width=8.0cm,angle=-90}
\label{g1t2}
\caption{Envelope of the first order correlation functions $G_X^{(1)}(\tau)$
(upper panel) and $G_C^{(1)}(\tau)$ (two lower panels)
at the stationary limit ($t \rightarrow \infty $). $g=1$, $\Delta=5$,
$\gamma=0.1$, $\kappa=5$ and $P=1$.}
\end{figure}
\begin{figure}[H]
\psfig{figure=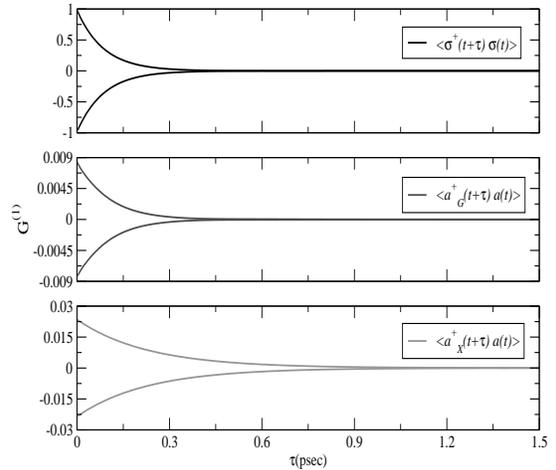,height=8.0cm,width=8.0cm,angle=-90}
\label{g1t3}
\caption{Envelope of the first order correlation functions $G_X^{(1)}(\tau)$
(upper panel) and $G_C^{(1)}(\tau)$ (two lower panels)
at the stationary limit ($t \rightarrow \infty $). $g=1$, $\Delta=5$,
$\gamma=0.1$, $\kappa=5$ and $P=15$.}
\end{figure}

The case of Fig. 13 corresponds to a high pumping dominating on the effect of
$\kappa $. As discussed above, one expects a strong damping of the oscillator with
frequency $\omega _X$. Therefore, only one peak at $\omega _X- \Delta$ is observed.
Moreover, the peak shown in this spectrum is significantly narrower than the
features observed in any other case we have analyzed as, for instance, the ones
in the following figures.

Fig. 14 shows the result with a high value of the rate emission $\kappa =5$ and a
smaller, although non negligible, pumping $P=1$. This case corresponds to 
$\overline{N_{ph}}=0.027$ and $g^{(2)}(0)=0.396$, i.e. to a sub-poissonian 
distribution of a small number of cavity photons. Now, the strong damping of 
the mode with frequency $\omega _X- \Delta$ reduces its intensity 
which becomes smaller than a second peak at $\omega _X$.
\begin{figure}
\psfig{figure=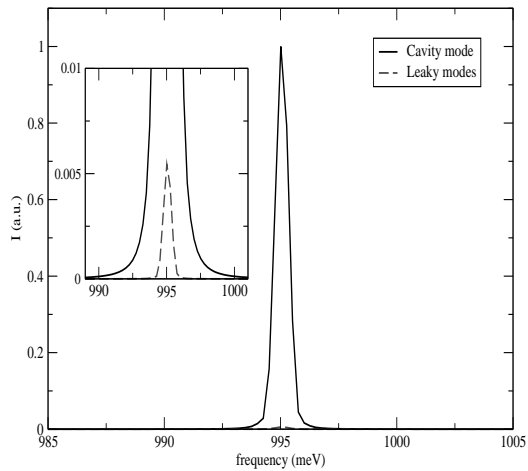,height=8.0cm,width=8.0cm,angle=-90}
\label{g1w1}
\caption{Spectrum of emission for $g=1$, $\omega _X=1000$, $\Delta=5$,
$\gamma=0.1$, $\kappa=0.1$ and $P=15$.}
\end{figure}

In the intermediate case shown in Fig. 15, the dephasing of the two modes is
similar and this produces a very wide peak at the spectrum as observed in the
figure.
\begin{figure}
\psfig{figure=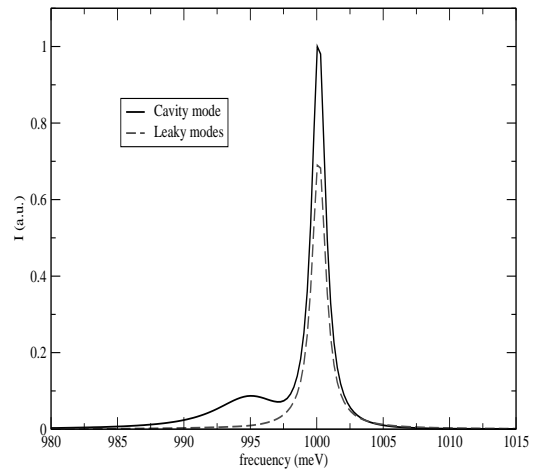,height=8.0cm,width=8.0cm,angle=-90}
\label{g1w2}
\caption{Spectrum of emission for $g=1$, $\omega _X=1000$, $\Delta=5$,
$\gamma=0.1$, $\kappa=5$ and $P=1$.}
\end{figure}

Let us finally analyze more carefully the case of high quality cavity (already
studied in Fig. 13). For this purpose, we maintain fixed a low value $\kappa =0.1$
and let the pumping vary in a wide range of values. Figs. 16 and 17 show the
spectra of the stimulated (cavity) and spontaneous (leaky) emissions respectively.
First of all one must observe that scale in Fig. 16 is more than 200 times 
larger than that of Fig. 17 as it could be expected from the results and
discussion of Fig. 13. The main result to be drawn from Figs. 16 and 17
is that the most intense emission corresponds to the range
$10 \leq P \leq 30 $ simply because this is the range of higher number of
photons inside the cavity as shown in Fig. 7.
\begin{figure}
\psfig{figure=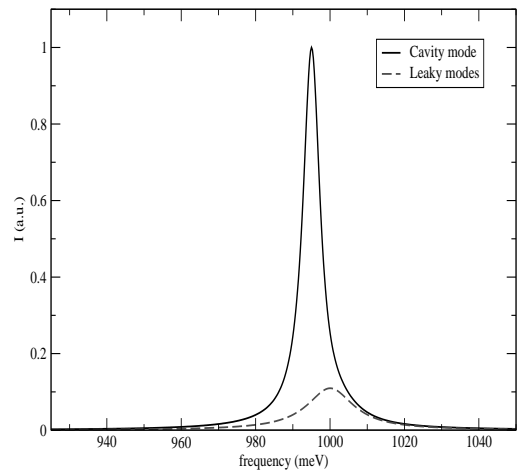,height=8.0cm,width=8.0cm,angle=-90}
\label{g1w3}
\caption{Spectrum of emission for $g=1$, $\omega _X=1000$, $\Delta=5$,
$\gamma=0.1$, $\kappa=5$ and $P=15$.}
\end{figure}

\begin{figure}
\includegraphics [clip,height=8cm,width=10.cm]{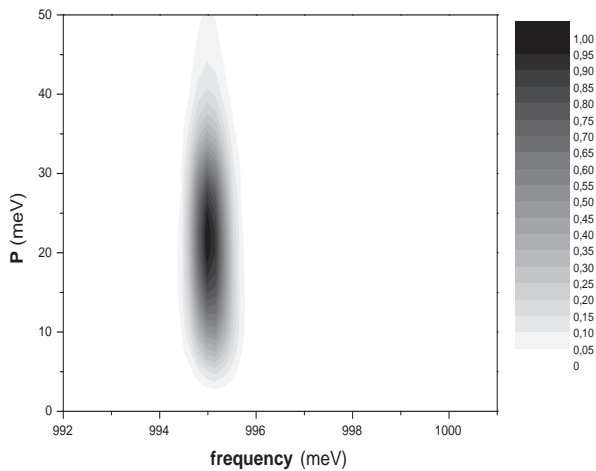}
\label{3Dg1w1}
\caption{Spectrum, in grey scale, of stimulated emission (cavity modes escaping from
the cavity) as a function of frequency and $P$. $g=1$, $\omega _X=1000$,
$\Delta=5$, $\gamma=0.1$ and $\kappa=0.1$.}
\end{figure}
\section{Summary}
We have described the dynamics of a QD embedded in a semiconductor microcavity
by means of a two-level system strongly coupled to a single cavity mode.
The system is continuously excited by incoherent pumping of excitons. 
Two different sources of photon emission are considered:
spontaneous emission through a leaky mode and stimulated emission of a cavity
mode escaping from the cavity.
The time evolution of first and second order coherence functions is calculated.
When pumping dominates over emission rates, a large number of cavity photons
can be stored in the cavity. Further increase of the pumping introduces dephasing
and a decrease of the number of cavity photons. These different regimes are also
characterized by poissonian or gaussian photon distributions inside the cavity.
Sub-poissonian distributions can be obtained for a range of parameters, in which the
pumping rate is very small, and the quantum nature of the QD-cavity 
system manifests itself in the emitted light.
Finally, we have studied the emission spectrum of our system.
In the case of high pumping dominating on the rate emission
$\kappa $, one gets a strong damping of the oscillator corresponding to 
the exciton level with frequency $\omega _X$, and only one very narrow peak 
at $\omega _X- \Delta$ is observed.
When the rate emission $\kappa $ is higher than a
non negligible pumping rate $P$, the strong damping of the mode
with frequency $\omega _X- \Delta$ allows to observe the two modes with a
higher intensity for the mode at $\omega _X$.
In a intermediate case, the dephasing of the two modes is
similar producing a very wide peak at the spectrum.
\begin{figure}[H]
\includegraphics [clip,height=8cm,width=10.cm]{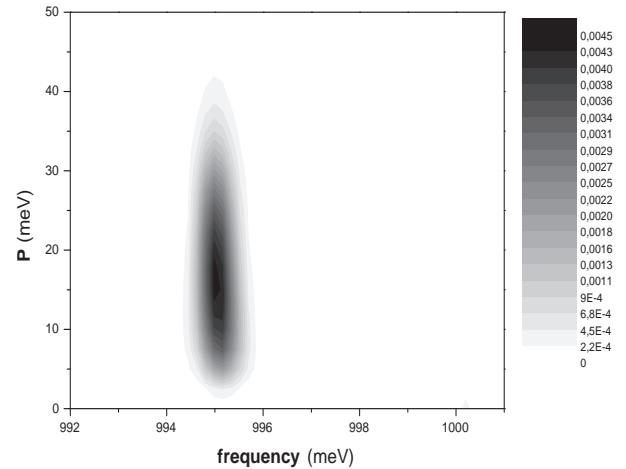}
\label{3Dg1w2}
\caption{Spectrum, in grey scale, of spontaneous emission (emission of leaky
modes) as a function of frequency and $P$.
$g=1$, $\omega _X=1000$, $\Delta=5$, $\gamma=0.1$ and $\kappa=0.1$.}
\end{figure}

\section{Acknowledgements}
We are indebted to Dr. P. Hawrylak for helpful discussions within the framework 
of the CERION2 project.
Work supported in part by MCYT of Spain under contract No. MAT2002-00139 and CAM
under Contract No. 07N/0042/2002. The numerical work has been carried out in
part at the CCC of UAM within the project QUANTUMDOT.


\end{document}